# Ferroelectric nematics: Materials with high permittivity or low resistivity?

Nataša Vaupotič[1,2,*], Tine Krajnc[1], Ewa Gorecka[3], Damian Pociecha[3], Vojko Matko[4]

[1] University of Maribor, Faculty of Natural Sciences and Mathematics, Koroška 160, 2000 Maribor, Slovenia
[2] Jozef Stefan Institute, Jamova 39, 1000 Ljubljana, Slovenia
[3] University of Warsaw, Faculty of Chemistry, Zwirki i Wigury 101, 02-089 Warsaw, Poland
[4] University of Maribor, Faculty of Electrical Engineering and Computer Science, Koroška 46, 2000 Maribor

* Corresponding author: natasa.vaupotic@um.si

**Abstract**

Two models have recently been proposed for a description of dielectric spectroscopy measurements of ferroelectric nematics ($N_F$) in thin planar capacitors. The polarization-external capacitance Goldstone reorientation mode (PCG model) considers the $N_F$ layer between the electrodes as an effective low resistivity material, the resistivity being inversely proportional to the square of polarisation magnitude. The "high-$\varepsilon$" model considers the $N_F$ material as having a huge permittivity due to the ease of polarisation rotation. In this paper we study implications of both models and show, why both models describe majority of the observed dielectric spectroscopy results equally well. We point out differences among the models' predictions and explain why some observations can be explained only by the high-$\varepsilon$ model. The major difference between the models is that the high-$\varepsilon$ model predicts that the increase in the cell thickness can lead to an increase in the frequency range within which capacitors filled with $N_F$ material can be used for energy storage while within the PCG model this frequency range reduces with increasing capacitor thickness. Within both models a crucial parameter which determines the behaviour of the capacitors filled with a $N_F$ material is parasitic resistance, primarily due to the electrode resistance. We present measurements of electrode resistance and find that in ITO cells it is of the order of few hundred ohms.

**Introduction**

Due to its polar nature and high spontaneous polarization, a ferroelectric liquid crystalline (LC) nematic phase ($N_F$) [1–8] is expected to have also a high value of the relative permittivity as characteristic for solid state ferroelectric materials. The values obtained by the dielectric spectroscopy (DS) measurements are indeed very high [2,9–15], but can vary by several order of magnitudes, depending on the type and thickness of the parallel plate capacitors filled with a $N_F$ material. This observations led to the proposal that the block reorientation of polarization in AC fields renders the $N_F$ layer effectively electrically conductive, which enables charging of very thin surface layers of surfactants [16]. Thus, what is measured by DS is, in fact, the capacitance of thin surface layers and the mode observed in DS is due to the charging/discharging of surface capacitors through the $N_F$ layer, which acts as a resistor with low ohmic resistance, its resistivity, $\rho_{LC}$, depending on the material rotational viscosity $\gamma$ and polarization $P$ as $\rho_{LC} = \gamma/P^2$. The observed mode is called the polarization-external capacitance Goldstone reorientation mode (PCG). The manuscript describing the PCG model was first published on arXiv already in 2022 [17], which motivated several research groups to additional studies to check whether the observed response is indeed a PCG mode or to design studies in such a way as to exclude the thin surface layer by using bare gold electrodes [11]. However, Clark et al. pointed out [16] that even in the case of bare electrodes there is a thin layer of LC molecules next to the electrodes that behaves differently than the bulk $N_F$ layer in the capacitor. This thin layer behaves in the same way as a thin layer of surfactant, as confirmed by recent studies [18–20]. Moreover, Erkoreka et al. [20] studied the properties of this thin layer of LC molecules in combination with the PCG model to find the dielectric tensor of material DIO, such that the contribution of the rotation of polarization is excluded from the tensor components.

Recently Adaka et al. [19] reported confirmation of the PCG model by studying DS results as a function of the cell thickness and thickness of the surface layer of surfactant. We, on the other hand, focused on an effect, which was neglected in the interpretation of data within the PCG model. According to the PCG model the low frequency capacitance of cells with bare electrodes should not depend on the cell thickness, as it is determined only by the thickness of the thin layer of LC material anchored to the electrode. However, experimentally it was observed that the low frequency capacitance for cells of different thickness differ [18]. We also observed that the ratio of capacitances for cells with varying thickness does not scale with their thickness ratio [18]. Similar behaviour was also observed for cells with thin surfactant layers ensuring homeotropic surface anchoring. For this reason, we proposed an alternative model to account for DS results [18], treating $N_F$ as a material with a high relative permittivity (high-$\varepsilon$ model). The $N_F$ material is not conductive, and for nonconductive materials their large response in external electric fields due to the reorientation of permanent dipoles is usually incorporated in the relative permittivity. However, as pointed out already within the studies of photovoltaic perovskites, the capacitance measurements are not always reliable in determining the relative permittivity, because capacitance of thin surface layers can be comparable to the capacitance of the bulk material. The photoinduced huge increase in relative permittivity observed in lead halide perovskite solar cells [21] was thus suggested to result from the change in the surface layer thickness [22] under illumination and not the increase in relative permittivity of the perovskite material.

In ref [18] we showed that DS results are well predicted by both models; however, the high-$\varepsilon$ model, in addition, explains also the difference in the low-frequency capacitance. We also showed that in the case of ITO electrodes, it is essential to consider the resistivity of electrodes. This is important especially for the PCG model, because the resistance of electrodes can be comparable to the assumed low resistance of the $N_F$ layer. Another point in favour of the high-$\varepsilon$ model is the fact that the rotational viscosity, predicted by the PCG model, is always much higher, up to three orders of magnitude, than viscosity measured by the switching time in electric field [18,19].

The fact that, up to the points raised above, both models explain well the results of DS measurements, set us to consider in more detail implications of both models. So, in this paper we first compare the models in respect to the capability of the $N_F$ capacitor to store electric charge. Then we compare the models regarding the complex impedance of the cell as a function of cell thickness, thickness of thin surface layers and resistance of electrodes. We also compare the apparent real and imaginary part of complex permittivity predicted by both models, which we calculate in the way usually done in the interpretation of the DS results of non-conductive materials. There, the current through the cell in phase ($I_0$) with voltage ($U$) across the cell is related to the imaginary part of complex permittivity ($\varepsilon''$) as

$$\varepsilon''_{AP} = \frac{I_0}{C_0 \omega U} \qquad (1)$$

and the current that is phase shifted by $\pi/2$ ($I_{\pi/2}$) to the real part of complex permittivity ($\varepsilon'$):

$$\varepsilon'_{AP} = \frac{I_{\pi/2}}{C_0 \omega U} \; , \qquad (2)$$

where $C_0$ is a capacitance of an empty capacitor (cell) and $\omega$ is the angular frequency of the AC voltage. The index "AP" standing for apparent, was added to the real and imaginary part of $\varepsilon$ in **eqs. (1)** and **(2)**, because in the case of highly polar material, such an interpretation of the DS results does not give the actual values of real ($\varepsilon'$) and imaginary ($\varepsilon''$) parts of complex permittivity. In the case of the PCG model these two values are not related to the relative permittivity of N$_F$ at all, while in the case of the high-$\varepsilon$ model they are related to $\varepsilon'$ and $\varepsilon''$, but in a rather complex way.

**Equivalent electric circuit of a capacitor filled with ferroelectric nematic**

The equivalent electric circuit for both models is essentially the same and is shown in **Figure 1**. The parallel plate capacitor filled with a LC material consists of electrodes, thin surface layers (surfactant or molecules of the studied material anchored at the surface) and a slab of liquid crystal. In the equivalent electric circuit, electrodes are presented by a resistor with resistance $R_e$. This resistance includes also parasitic resistance due to wiring. Thin surface layers are presented each by a parallel connection of a capacitor with capacitance $C_S$ and resistor with resistance $R_S$. Usually, surface layers can be treated as having an infinite resistance. Their finite resistance is important only at very low frequencies, where $R_S \leq (\omega C_S)^{-1}$. The capacitance $C_S$ depends on the dielectric constant $\varepsilon_S$ of the material that forms a thin surface layer and thickness of the surface layer ($d_S$). We express $C_S$ in terms of capacitance of an empty capacitor as $C_S = \varepsilon_S C_0 d/d_S$, where $C_0 = \varepsilon_0 S/d$, $\varepsilon_0$ is the permittivity of vacuum, $d$ the capacitor (cell) thickness and $S$ surface area of electrodes. For $\varepsilon_S$ we assume that it is real and frequency independent, which is a usual assumption at frequencies up to few MHz. The LC layer is presented by a parallel connection of a capacitor with capacitance $C_{LC}$ and resistor with resistance $R_{LC}$. The crucial difference between the PCG model and high-$\varepsilon$ model is the way of treating $C_{LC}$ and $R_{LC}$. Within the PCG model, $C_{LC}$ is irrelevant in the interpretation of DS measurements, because $R_{LC}$ is so low that the current flows only through the resistor $R_{LC}$. As explained above, in the PCG model the N$_F$ layer is treated as an effective conductor due to the block-reorientation of polarization. In a parallel plate capacitor $R_{LC} = \rho_{LC} d S^{-1} = \gamma d P^{-2} S^{-1}$. On the other hand, in the high-$\varepsilon$ model, a finite resistance $R_{LC}$ is due to ionic impurities. The resistivity of LCs is of the order of $10^8$ $\Omega$m or even larger and can be assumed to be frequency independent within the typical measurement range of DS. With $R_{LC} \gg (\omega C_{LC})^{-1}$, the capacitor representing the N$_F$ layer ($C_{LC}$) becomes a crucial element in the equivalent electric circuit. Its capacitance $C_{LC} = C_0 \varepsilon(\omega)$ is expressed with a complex relative permittivity of the LC material $\varepsilon(\omega) = \varepsilon' - i\varepsilon''$, which depends on the angular frequency $\omega$. We approximate complex relative permittivity by the Debye relaxation model:

$$\varepsilon(\omega) = \varepsilon_\infty + \frac{\varepsilon_{lf} - \varepsilon_\infty}{1 - i\omega\tau_D} \; ,$$

where $\varepsilon_\infty$ is a high-frequency and $\varepsilon_{lf}$ low-frequency relative permittivity and $\tau_D$ is the Debye relaxation time. The real and imaginary part of complex relative permittivity are thus

$$\varepsilon' = \varepsilon_\infty + \frac{\varepsilon_{lf} - \varepsilon_\infty}{1 + \omega^2 \tau_D^2} \tag{3}$$

and

$$\varepsilon'' = \omega\tau \frac{\varepsilon_{lf} - \varepsilon_\infty}{1 + \omega^2 \tau_D^2}, \tag{4}$$

respectively. The frequency dependence of $\varepsilon'$ (**eq. (3)**) and $\varepsilon''$ (**eq. (4)**) is well known and is shown in **Figure 1d**. The value of $\varepsilon''$ is maximum at

$$\omega_0 = \frac{1}{\tau_D}$$

and equals to $(\varepsilon_{lf} - \varepsilon_\infty)/2$. If $\varepsilon_{lf} \gg \varepsilon_\infty$, then the maximum value of $\varepsilon''$ equals to half the value of the low-frequency relative permittivity. The Debye relaxation model could also be used for the material of the thin surface layer, but this relaxation frequency is assumed to be above the typical frequency range of DS measurements.

The complex impedance of the equivalent electric circuit shown in **Figure 1** is

$$Z_c = R_e + \left(\frac{1}{2R_S} + \frac{i\omega C_S}{2}\right)^{-1} + \left(\frac{1}{R_{LC}} + i\omega C_{LC}\right)^{-1}. \tag{5}$$

Let us consider some specific cases of the above expression (**eq. (5)**). First (**Case 1**) we assume that $R_{LC}$ and $R_S$ are very large, $R_e \approx 0$ and $C_S \gg C_{LC}$ (i.e. $\varepsilon'$ is low). The complex impedance $Z_c$ reduces to

$$Z_c = \frac{1}{i\omega C_{LC}} = \frac{1}{i\omega C_0(\varepsilon' - i\varepsilon'')}.$$

When an AC voltage with amplitude $U_0$ is applied across the cell, $U = U_0 e^{i\omega t}$, the current ($I$) through the cell is $I = U Z_c^{-1}$:

$$I = \omega C_0 U(\varepsilon'' + i\varepsilon').$$

The current in phase with the voltage is thus directly proportional to the imaginary part and the current that is phase shifted by $\pi/2$ is proportional to the real part of the relative permittivity of the LC material. By using **eqs. (1)** and **(2)** we calculate $\varepsilon'$ and $\varepsilon''$; their frequency dependence is shown in **Figure 1d**. The maximum value of $\varepsilon''$ is at $\omega_0 = 1/\tau_D$.

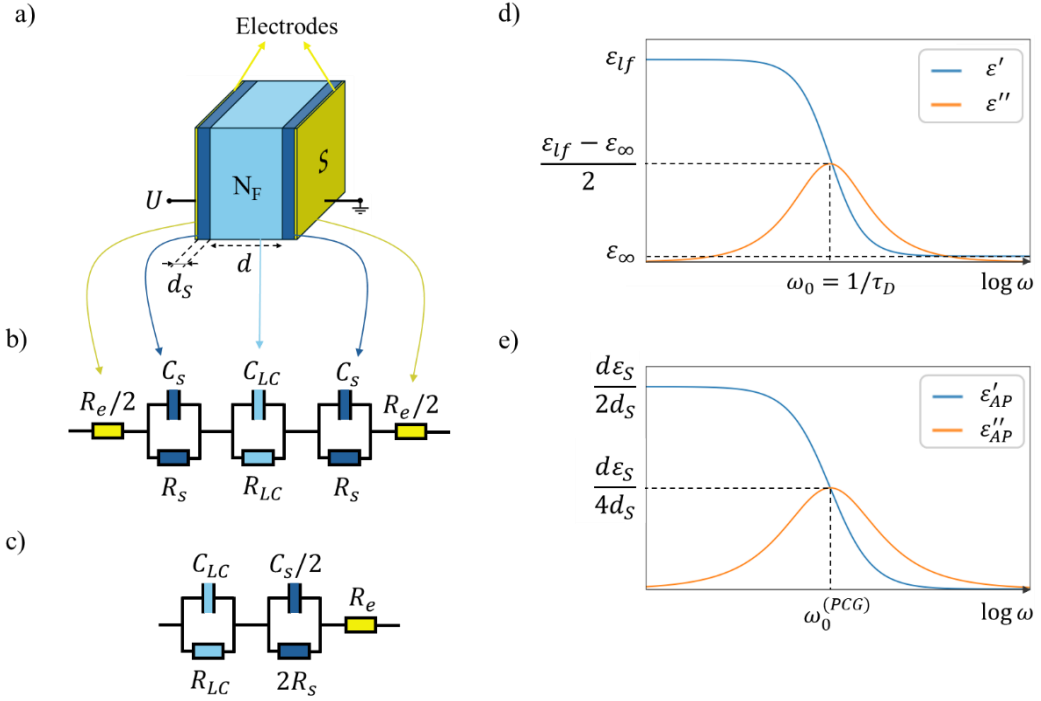

**Figure 1.** a) Liquid crystal cell consists of electrodes, thin surface layer of thickness $d_S$ and a bulk ferroelectric nematic phase (N$_F$) of thickness $d$. b) Equivalent electric circuit of the cell shown in a). $C_S$ is a capacitance of a surface layer and $R_S$ its resistance, $R_e$ is a resistance of both electrodes, containing also parasitic resistance due to wiring. $C_{LC}$ is a frequency dependent capacitance of the N$_F$ layer and $R_{LC}$ its resistance. c) The same as b) but with combined surface capacitors and resistances of electrodes. d) Real and Imaginary part of the complex permittivity according to the Debye model; $\varepsilon_\infty$ is a high-frequency and $\varepsilon_{lf}$ low-frequency relative permittivity and $\tau_D$ is the Debye relaxation time. The maximum of $\varepsilon''$ is at angular frequency $\omega_0$. e) Apparent values of the real ($\varepsilon'_{AP}$) and imaginary ($\varepsilon''_{AP}$) part of complex permittivity for the PCG model; $\omega_0^{(PCG)}$ is the frequency given by **eq. (6)**.

Next (**Case 2**), we make an unphysical assumption that $C_S \ll C_{LC}$, but keep $R_{LC}$ and $R_S$ large and $R_e \approx 0$. The complex impedance (**eq. (5)**) now reduces to

$$Z_c = \frac{2}{i\omega C_S}.$$

In this case we measure only the current that is phase shifted to voltage by $\pi/2$: $I = iU\omega C_S/2$. This current does not depend on relative permittivity of the LC material at all. By using **eq. (2)**, we thus calculate some apparent value of the relative permittivity, which has no relation to the material properties, instead, it is related to the cell thickness, thickness of the surface layer and relative permittivity of the surface layer as $\varepsilon'_{AP} = C_S/(2C_0) = \varepsilon_S d/(2d_S)$.

In **Case 3** we assume that $R_S$ is large while $R_{LC}$ is so low, that $C_{LC}$ can be neglected (PCG model). If $R_e$ is not negligibly small compared to $R_{LC}$, then

$$Z_c = R_e + R_{LC} + \frac{2}{i\omega C_S}.$$

In this case
$$I = \frac{U}{(R_e + R_{LC})^2 + \frac{4}{\omega^2 C_S^2}} \left( R_e + R_{LC} + \frac{2i}{\omega C_S} \right).$$

The apparent real and imaginary parts of complex permittivity are calculated by **eqs. (1)** and **(2)** and the low-frequency value of $\varepsilon'_{AP}$ is found to be

$$\varepsilon'_{AP}(\omega \to 0) = \frac{C_S}{2C_0} = \frac{d}{2d_S} \varepsilon_S, \qquad (6)$$

which is the same as in Case 2, however unlike in Case 2, $\varepsilon'_{AP}$ is now frequency dependent. The peak value of the imaginary part of apparent relative permittivity ($\varepsilon''_{AP,0}$) is at an angular frequency at which the real and imaginary part of $Z$ are equal: $\omega_0 = 2C_S^{-1}(R_e + R_{LC})^{-1}$, which can be expressed as

$$\omega_0 = \frac{2d_S}{\varepsilon_S \varepsilon_0 S \left( R_e + \frac{\rho_{LC} d}{S} \right)}. \qquad (7)$$

If $R_e \ll \rho_{LC} d/S$, then $\omega_0 \propto d_S/d$. It is straightforward to show that at $\varepsilon''_{AP,0} = \varepsilon'_{AP}(\omega \to 0)/2$.

Finally (**Case 4**), if $R_{LC}$ and $R_S$ are large, but no other assumptions are made, then

$$Z_c = R_e + \frac{2}{i\omega C_S} + \frac{1}{i\omega C_{LC}}.$$

By using $C_{LC} = C_0(\varepsilon' - i\varepsilon'')$ we can find the real ($Z_r$) and imaginary ($Z_i$) part of $Z_c = Z_r + iZ_i$:

$$Z_r = R_e + \frac{\varepsilon''}{\omega C_0 (\varepsilon'^2 + \varepsilon''^2)}$$

and

$$Z_i = -\frac{2}{\omega C_S} - \frac{\varepsilon'}{\omega C_0 (\varepsilon'^2 + \varepsilon''^2)}.$$

The current through the cell is

$$I = \frac{U}{|Z_c|^2} (Z_r - iZ_i).$$

By using **eqs. (1)** and **(2)**, we calculate the apparent values of the real and imaginary part of complex relative permittivity:

$$\varepsilon'_{AP} = -\frac{Z_i}{|Z_c|^2 \omega C_0}$$

and

$$\varepsilon''_{AP} = \frac{Z_r}{|Z_c|^2 \omega C_0}. \qquad (8)$$

In this case the apparent values $\varepsilon'_{AP}$ and $\varepsilon''_{AP}$ depend on $\varepsilon'$, $\varepsilon''$, $C_0$ and $C_S$. The low-frequency value of $\varepsilon'_{AP}$ is found to be

$$\varepsilon'_{AP}(\omega \to 0) = \frac{C_S \varepsilon_{lf}}{C_S + 2C_0 \varepsilon_{lf}} = \frac{\varepsilon_{lf}}{1 + 2\frac{d_S}{d}\frac{\varepsilon_{lf}}{\varepsilon_S}} , \qquad (9)$$

which reduces to $\varepsilon'_{AP} = \varepsilon_{lf}$, i.e. to the actual material value, if $\varepsilon_{lf} \ll C_S/C_0$, meaning $\varepsilon_{lf} \ll d/d_S$. If, on the other hand, $\varepsilon_{lf} \gg d/d_S$, expression in **eq. (9)** reduces to the value predicted by the PCG model (**eq. (6)**).

The value of $\varepsilon''_{AP}$ will have maximum at the angular frequency $\omega_0$ at which the real and imaginary part of the complex impedance are the same: $Z_r = Z_i$. In general, $\omega_0$ can be found only numerically. However, if one considers the case of very large $R_{LC}$ and $R_S$, $\varepsilon_{lf} \gg \varepsilon_\infty$ and $C_S \gg C_0 \varepsilon_\infty$ then

$$\omega_0 = \frac{1 + 2\frac{d_S}{d}\frac{\varepsilon_{lf}}{\varepsilon_S}}{\tau_D + C_0 R_e \varepsilon_{lf}} . \qquad (10)$$

The ratio $d_S/d$ is of the order of $10^{-3}$ or lower. When $\varepsilon_{lf}$ is low, the expression for $\omega_0$ again reduces to $\omega_0 = 1/\tau_D$, thus the relaxation time of material is directly measured through the frequency of maximum $\varepsilon''_{AP}$. But if the low-frequency relative permittivity $\varepsilon_{lf}$ is of the order of $10^3$ or larger, then both the thickness of the cell and thickness of thin surface layers will have a strong effect on the position of the peak value of $\varepsilon''_{AP}$, shifting it to higher frequencies. On the other hand, when $R_e \sim \tau_D/(C_0 \varepsilon_{lf})$, the resistance of electrodes affects the position of the peak value and shifts it to lower frequencies.

The frequency dependence of $\varepsilon'_{AP}$ and $\varepsilon''_{AP}$ for the high-$\varepsilon$ model resembles the frequency dependence of the real and imaginary part of complex permittivity resulting from the Debye model. So, **Figure 1d** is applicable, but the low-frequency value of the real part is given by **eq. (9)**, while the maximum of $\varepsilon''_{AP}$ is at angular frequency given by **eq. (10)** and equals to approximately half the low frequency value of $\varepsilon'_{AP}$.

**Capacity to store energy**

In this section, we consider the charge being stored on electrodes of a capacitor filled with a LC in the ferroelectric nematic phase. When considering the PCG model we can neglect capacitance of the N$_F$ layer because $R_{LC}$ is low. So, the potential drop will be on surface capacitors, $U_0/2$ on each capacitor, if we neglect a small current due to the finite resistance of surface layers. Thus, by filling the cell by N$_F$, we obtain a series connection of two nano-thin capacitors and their joint capacitance is

$$C = \frac{C_S}{2} = C_0 \frac{d}{2d_S}\varepsilon_S . \qquad (11)$$

Because, usually, $d_S \sim 10^{-3}d$, we have thus increased the capacitance by 3 orders of magnitude. The capacitor will effectively store energy at AC field frequencies up to $\omega_0$ given by **eq. (7)**. This frequency decreases with increasing cell thickness, while $C$ is independent of the cell thickness.

Within the high-ε model, all three capacitors in the equivalent electric circuit are charged. Their net capacitance is:

$$C = \left(\frac{2}{C_S} + \frac{1}{C_{LC}}\right)^{-1}.$$

At low frequencies, where $C_{LC} = C_0 \varepsilon_{lf}$, we have

$$C = C_0 \frac{\varepsilon_{lf}}{2\frac{d_S}{d}\frac{\varepsilon_{lf}}{\varepsilon_S}+1} \quad . \tag{12}$$

In the case of $\varepsilon_{lf}$ being very high, the expression for the capacitance $C$ from **eq. (12)** reduces to the capacitance obtained by the PCG model (**eq. (11)**). By a proper selection of $\rho_{LC}$ in the PCG model and $\tau_D$ in the high-$\varepsilon$ model, $\omega_0$ obtained from the two models is also similar. As a result, both models predict approximately the same energy storage capability. Note, that the factor next to $C_0$ in **eqs. (11)** and **(12)** is, in fact, the low-frequency value of $\varepsilon'_{AP}$ obtained for both models, see **eqs. (6)** and **(9)**. Thus, even though the low-frequency value of $\varepsilon'_{AP}$ obtained from DS is not the material value $\varepsilon_{lf}$, it is a value that gives a direct information on how much more charge can be stored on the capacitor plates, when the capacitor is filled with a LC material in the ferroelectric nematic phase, compared to the empty capacitor. The frequency $\nu_0 = \omega_0/(2\pi)$ at which $\varepsilon''_{AP}$ is maximum is the frequency up to which the capacitor can be used as a storage device. Because $\nu_0$ can be rather low, even of the order of few 10 Hz [11,18], the storage capability can be severely limited as already pointed out in [16]. Within both models the angular frequency $\omega_0$ is increased if $d_S$ is increased (see **eqs. (7)** and **(10)**), but increase in $d_S$ will reduce the capacitance (**eqs. (11)** and **(12)**).

However, there are two significant differences between the predictions of the two models. By comparing **eqs. (11)** and **(12)** we first observe that within the PCG model the capacitance $C$ depends on the thickness of the surface layer but not on the thickness of the cell, while the high-$\varepsilon$ model predicts that $C$ depends on $d$, unless $d_S \varepsilon_{lf} \gg d \varepsilon_S$.

The second difference is the following. The PCG model predicts that $\omega_0 \to 0$ when the cell thickness increases, so thick capacitors will not be useful to store energy at AC fields. On the other hand, the high-$\varepsilon$ model predicts (see **eq. (10)**) that at $d\varepsilon_S \gg d_S\varepsilon_{lf}$ and $\tau_D \to 0$, the angular frequency $\omega_0 \to (C_0 R_e \varepsilon_{lf})^{-1}$. In this case, the frequency up to which the capacitor will effectively store electric charge depends on the resistance of electrodes, it is independent of the thickness of the thin surface layers, and it increases linearly with the cell thickness (as opposed to the PCG model where $\omega_0 \propto d^{-1}$). If we take $C_0 = 100$ pF, $R_e = 100$ Ω and $\varepsilon_{lf} = 10^4$, we find $\omega_0 = 10^6$ s$^{-1}$, i.e. $\nu_0 = 160$ kHz, while $\varepsilon'_{AP}(\omega \to 0) \approx \varepsilon_{lf}$.

**Model predictions**

In this section we present some predictions of the high-$\varepsilon$ model and compare it to the predictions of the PCG model.

For the high-$\varepsilon$ model, the dependence of the maximum value of the apparent imaginary part of complex permittivity ($\varepsilon''_{AP,0}$) and the frequency of the peak ($\nu_0$) on the cell thickness ($d$) and on thickness of the surface layer ($d_S$) is given in **Figure 2**. At large values of cell thickness, $\varepsilon''_{AP,0}$ approaches the value $(\varepsilon_{lf} - \varepsilon_\infty)/2$ and $\nu_0$ goes to the Debye relaxation frequency $\nu_D = (2\pi\tau_D)^{-1}$. This is expected because by increasing the cell thickness, the capacitance of the LC layer decreases. When $C_{LC} \ll C_S$, only the LC layer capacitor is charged, and LC properties are directly obtained from DS measurements.

On the other hand, $C_{LC} \ll C_S$ is satisfied if $d_S$ is sufficiently low. At low $d_S$ we have $\varepsilon''_{AP,0} \approx (\varepsilon_{lf} - \varepsilon_\infty)/2$ and $\nu_0 \approx \nu_D$. The value of $\varepsilon''_{AP,0}$ decreases with increasing $d_S$. At large $d_S$, where the condition $C_{LC} \gg C_S$ is satisfied, only surface capacitors are charged. The frequency $\nu_0$ increases linearly with $d_S$ as given by **eq. (10)**.

The dependences $\varepsilon''_{AP,0}(\omega)$ given in **Figures 2a,c** are calculated for the high-$\varepsilon$ model without any simplifications of the equivalent circuit. This means that $Z_c$ given by **eq. (5)** was used to calculate $\varepsilon''_{AP}$ given by **eq. (8)** and its maximum was found. In such a way $\omega_0$ is found as well; however, in **Figures 2b,d** we plot $\omega_0$ as given by **eq. (10)** (i.e. $C_S \gg C_0 \varepsilon_\infty$ is assumed). This turns out to be a satisfactory approximation as shown in **Figure 3**, where we give $\nu_0$ as a function of $d$ and $d_S$ for both calculations, with **(eq. (10))** and without simplifications. At a chosen set of parameters, the error is negligibly small (less than 1 %), so the approximation can be used. By decreasing $\varepsilon_{lf}$ and/or increasing $d_S$ the relative error increases. For example, if $d_S$ is increased to $d_S = 5$ nm and $\varepsilon_{lf}$ reduced to $\varepsilon_{lf} = 10^4$, the rest of parameters being the same as in **Figure 3**, the error is of the order of 1 % at $d = 10$ $\mu$m and increases to approximately 2 % at $d = 1$ $\mu$m.

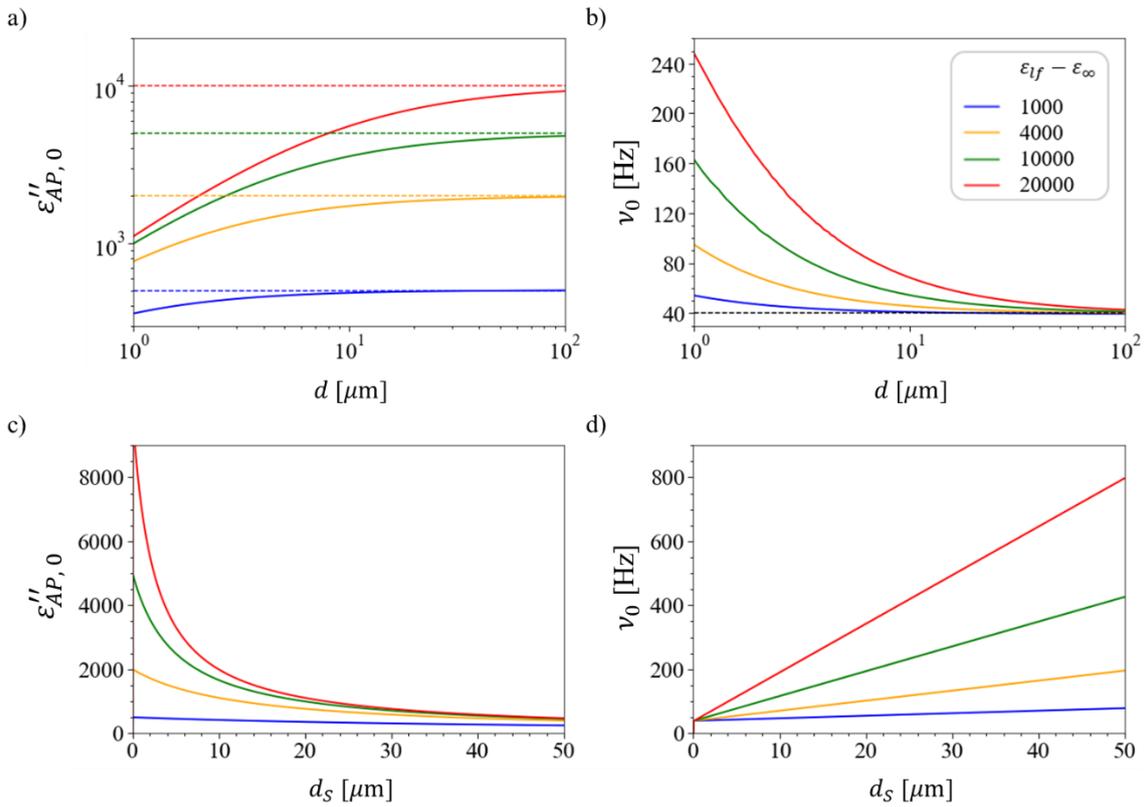

**Figure 2.** *High-$\varepsilon$ model.* a) Maximum (peak) value of the apparent imaginary part of complex permittivity $(\varepsilon''_{AP,0})$ and b) the frequency $(\nu_0)$ of the peak as a function of cell thickness ($d$) at different values of $\varepsilon_{lf} - \varepsilon_\infty$; c) $\varepsilon''_{AP,0}$ and d) $\nu_0$ as a function of surface layer thickness ($d_S$) at different values of $\varepsilon_{lf} - \varepsilon_\infty$. Dotted lines in a) present $(\varepsilon_{lf} - \varepsilon_\infty)/2$, dotted line in b) presents the Debye frequency $\nu_D = 1/(2\pi\tau_D)$ for $\tau_D = 4$ ms. Parameter values: $R_S = 1$ M$\Omega$, $\rho_{LC} = 10^8$ $\Omega$m, $R_e = 100$ $\Omega$, $S = 100$ mm$^2$, $\varepsilon_S = 10$, $d_S = 2$ nm in a), b) and $d = 10$ $\mu$m in c) and d).

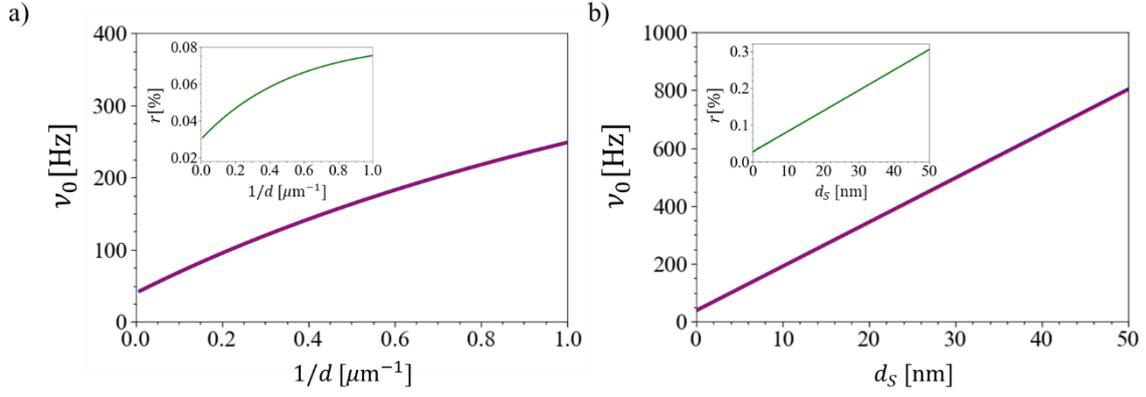

Figure 3. *High-ε model.* Frequency ($\nu_0$) of the peak value of $\varepsilon''_{AP}$ as a function of a) cell thickness ($d$) and b) thickness of the surface layer ($d_S$), plotted by using the approximation given in **eq. (10)** and without the approximation. The insets give the relative difference ($r$) between the two values. Parameter values: $\varepsilon_{lf} - \varepsilon_\infty = 2 \cdot 10^4$, $\tau_D = 4$ ms, $R_S = 1$ MΩ, $\rho_{LC} = 10^8$ Ωm, $R_e = 100$ Ω, $S = 100$ mm², $\varepsilon_S = 10$, $d_S = 2$ nm in a) and $d = 10$ μm in b).

Because **eq. (10)** predicts linear dependence of $\nu_0$ on both, $d_S$ and $1/d$, if the term $C_0 R_e \varepsilon_{lf}$ in the denominator of **eq. (10)** is negligibly small compared to $\tau_D$, we give $\nu_0(d_S)$ and $\nu_0(1/d)$ in **Figure 4** at different values of parasitic resistance $R_e$ and at two values of the Debye relaxation time. We observe that, as soon as $R_e \neq 0$, the dependence $\nu_0(1/d)$ deviates from being linear because when $d$ decreases, $C_0$ increases, thus the importance of the second term in the denominator on the right-hand side of **eq. (10)** increases. At low enough $\tau_D$ and in thin cells, it can happen that the right-hand side terms in the nominator and denominator of **eq. (10)** prevail, thus $\omega_0 \approx 2d_S(\varepsilon_S \varepsilon_0 S R_e)^{-1}$ is determined by the electrode resistance and thickness of the surface layer. In such a case (shown in **Figure 4c**) the frequency $\nu_0$ can even decrease with decreasing $d$ (i.e. increasing $1/d$). The PCG model also predicts a linear dependence of $\nu_0$ on $d_S$ and a linear dependence of $\nu_0$ on $1/d$, if $R_e$ is negligibly small. In **Figure 4**, the dependencies following from the PCG model are given by dashed lines. The parameters are chosen such, that the frequencies $\nu_0$ are of the same order of magnitude within both models.

From **Figure 4** one can already observe that the difference between the frequencies predicted by both models decreases with decreasing $d$ and increasing $d_S$. If frequencies $\nu_0$ measured by DS are rather high as in the case of material studied in [19], the measurements can be equally described by both models as shown in **Figure 5**. The parameters can be chosen such that the lines in **Figure 5b** coincide, but to show the lines for both models, we chose such parameters that there is a slight difference in the predicted values. The cells used in [19] are quite atypical, because for the purpose of the study of the effect of the thickness of the surface layer, $d_S$ was rather large, from 90 nm up to 334 nm in cells with $d \approx 10$ μm. In cells with bare electrodes, where a thin layer of LC close to the electrodes behaves in the same way as a surfactant, $d_S$ is only few nanometres. At low $d_S$ the difference between the two models is evident, as shown in **Figure 5a**.

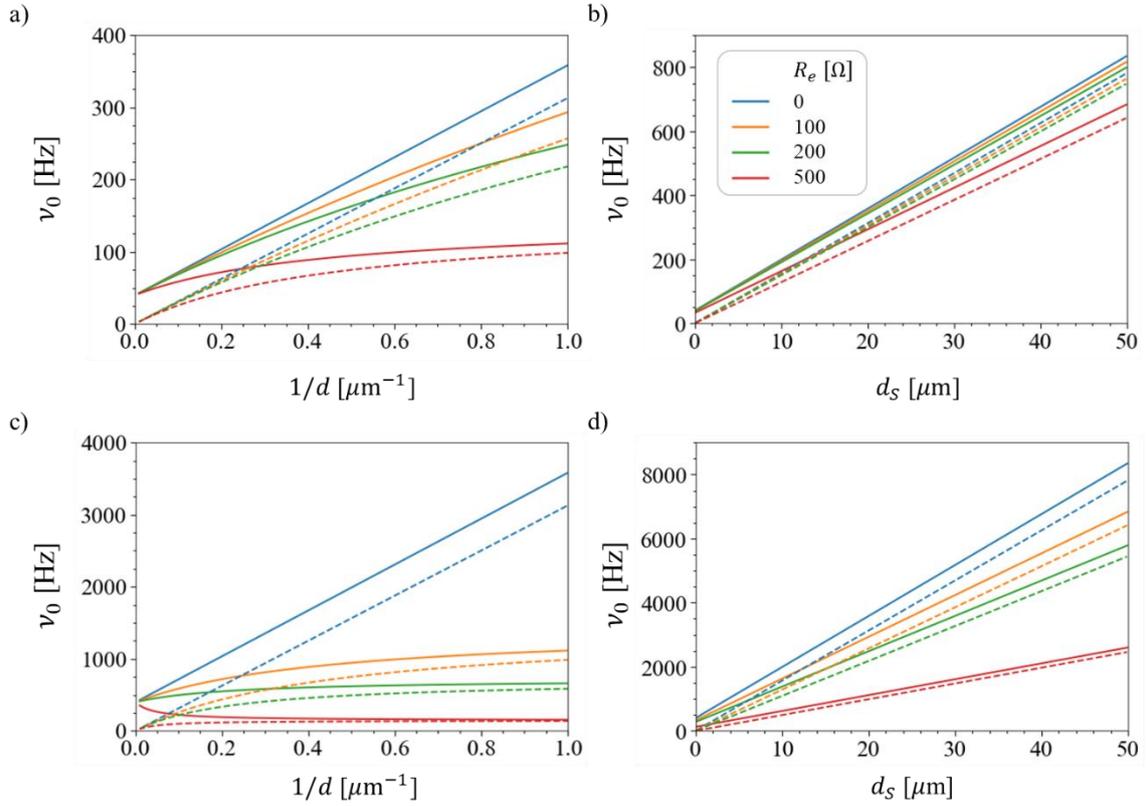

**Figure 4.** *High-ε model (solid lines) and PCG model (dashed lines).* Frequency ($\nu_0$) of the peak value of $\varepsilon''_{AP}$ as a function of a) cell thickness ($d$) and b) thickness of the surface layer ($d_S$), at $\tau_D = 4$ ms and c) $\nu_0(1/d)$ and d) $\nu_0(d_S)$ at $\tau_D = 0.4$ ms, for different values of the parasitic resistance $R_e$. Parameter values: $\varepsilon_{lf} - \varepsilon_\infty = 2 \cdot 10^4$, $R_S = 1$ MΩ, $S = 100$ mm², $\varepsilon_S = 10$, $\rho_{LC} = 10^8$ Ωm for the high-ε model; $\rho_{LC} = 23$ kΩm (a,b) and $\rho_{LC} = 2.3$ kΩm (c,d) for the PCG model; $d_S = 2$ nm in a) and c) and $d = 10$ μm in b) and d).

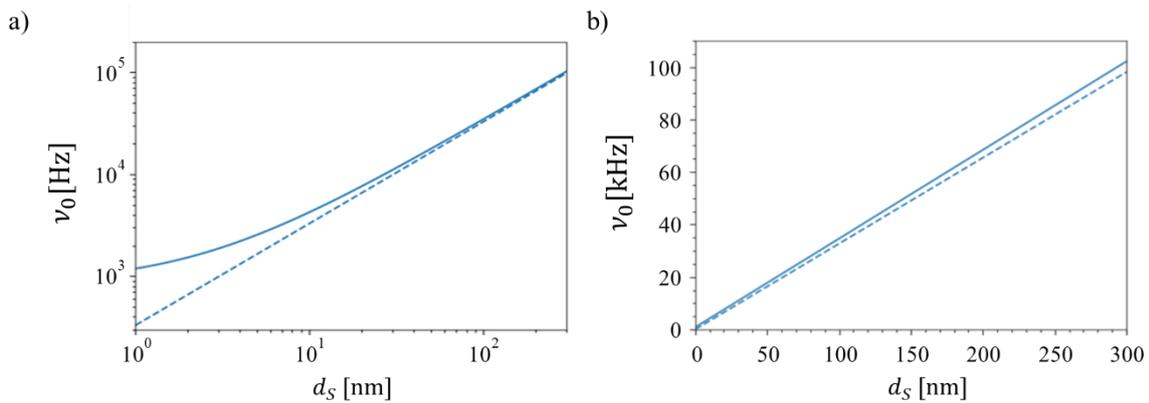

**Figure 5.** Frequency of the peak value of $\varepsilon''_{AP}$ as a function of the thickness of the surface layer ($d_S$) for the high-ε (solid lines) and PCG model (dashed lines) given in a a) logarithmic and b) linear scale. Parameter values: $\varepsilon_{lf} - \varepsilon_\infty = 2 \cdot 10^4$ for the high-ε model and $\rho_{LC} = 600$ Ωm for the PCG model, $R_e = 100$ Ω, $S = 50$ mm², $\varepsilon_S = 10$, $d = 10$ μm and $\tau_D = 0.1$ ms.

It is instructive to look in more detail at the dependence of the maximum value $\varepsilon''_{AP,0}$ and its position (frequency $\nu_0$) on the magnitude of $\varepsilon_{lf} - \varepsilon_\infty$ (**Figure 6**). We see that the value of $\varepsilon''_{AP,0}$ equals half the value of $\varepsilon_{lf} - \varepsilon_\infty$ only in some region of values $\varepsilon_{lf} - \varepsilon_\infty$. At low values of $\varepsilon_{lf} - \varepsilon_\infty$, $\varepsilon''_{AP,0}$ is higher than $(\varepsilon_{lf} - \varepsilon_\infty)/2$ while at large values of $\varepsilon_{lf} - \varepsilon_\infty$ it is lower. The physical reasons are very different. At low values of $\varepsilon_{lf} - \varepsilon_\infty$, one would expect that the actual material complex permittivity is measured by DS, but if $\nu_D$ is low, the peak of $\varepsilon''_{AP,0}$ is lost in the low-frequency increase of $\varepsilon''_{AP}$ due to the ion conductivity. At large values of $\varepsilon_{lf} - \varepsilon_\infty$, the peak value of $\varepsilon''_{AP}$ is lower than $(\varepsilon_{lf} - \varepsilon_\infty)/2$ because all three capacitors in the equivalent electric circuit of the cell are charged. From **Figure 6b** we see that by increasing $\varepsilon_{lf} - \varepsilon_\infty$, the frequency at which $\varepsilon''_{AP}$ is maximum shifts to frequencies higher than $\nu_D$. At low values of $\varepsilon_{lf} - \varepsilon_\infty$ one would expect $\nu_0 = \nu_D$. However, the contribution to $\varepsilon''_{AP}$ due to ionic conductivity, shifts this frequency to values lower than $\nu_D$ and at low enough frequency the peak disappears. If $R_S$ and/or $\rho_{LC}$ are assumed to be very high, then $\nu_0 \approx \nu_D$ at low frequencies, as expected (see **Figure 6b**). In this case, the value $\varepsilon''_{AP,0}$ also decreases linearly with decreasing $\varepsilon_{lf} - \varepsilon_\infty$ (see the inset to **Figure 6a**).

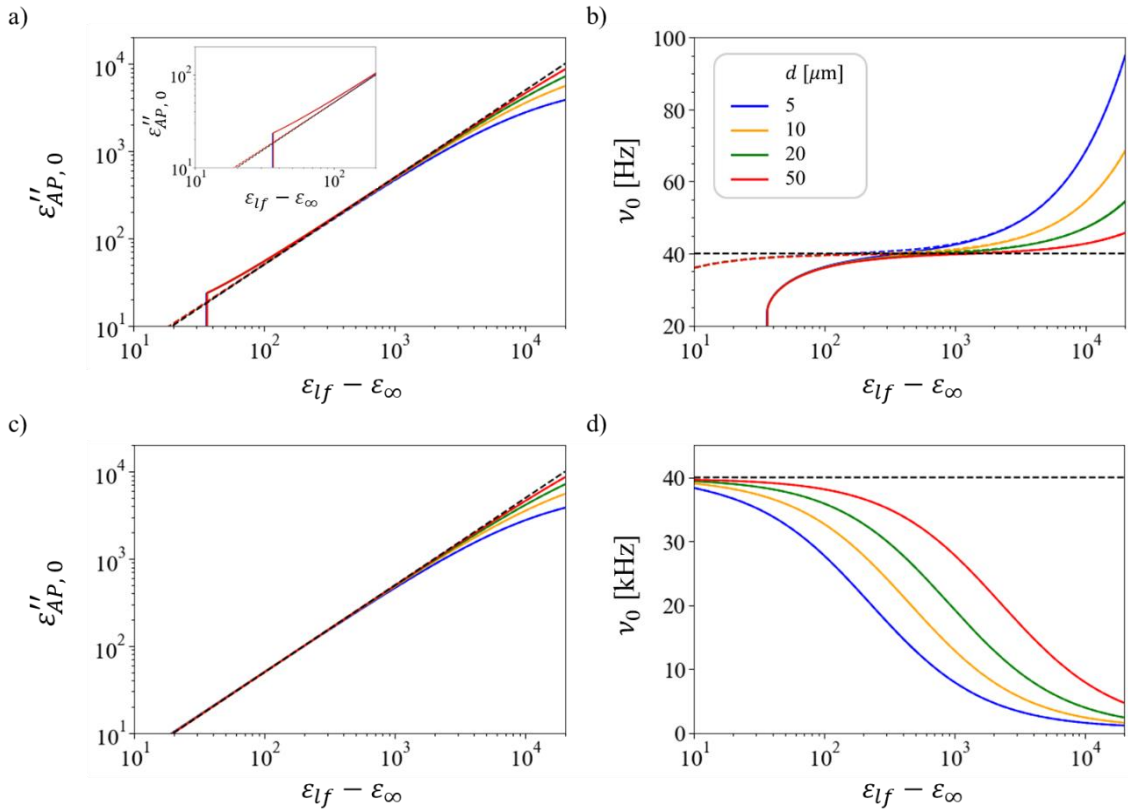

**Figure 6**. *High-ε model.* a) $\varepsilon''_{AP,0}$ and b) $\nu_0$ as a function of $\varepsilon_{lf} - \varepsilon_\infty$ at different values of $d$ at $\nu_D = 4$ ms and c) $\varepsilon''_{AP,0}$ and d) $\nu_0$ as a function of $\varepsilon_{lf} - \varepsilon_\infty$ at different values of $d$ at $\nu_D = 4$ μs. Black dashed lines in a) and c) present $(\varepsilon_{lf} - \varepsilon_\infty)/2$, and the Debye frequency $\nu_D = 1/(2\pi\tau_D)$ in b) and d). Parameter values: $R_S = 1$ MΩ, $\rho_{LC} = 10^8$ Ωm (solid coloured lines), $\rho_{LC} = 10^9$ Ωm (dashed coloured lines), $R_e = 100$ Ω, $S = 100$ mm$^2$, $\varepsilon_S = 10$, $d_S = 2$ nm.

Next, we look at the dependencies of $\varepsilon''_{AP,0}$ and $\nu_0$ on $\varepsilon_{lf} - \varepsilon_\infty$ at lower Debye relaxation times and thus higher Debye frequencies. In **Figure 6c,d** these dependencies are shown at the Deby relaxation time that is three orders of magnitude lower than in **Figure 6a,b**. In this case the Debye frequency is so high that $\varepsilon''_{AP,0} = (\varepsilon_{lf} - \varepsilon_\infty)/2$ at low enough values of $\varepsilon_{lf} - \varepsilon_\infty$, while at high values of $\varepsilon_{lf} - \varepsilon_\infty$ we again observe the departure from the linear dependence. On the other hand, the frequency $\nu_0$ reduces from the Debye relaxation frequency, detected at low $\varepsilon_{lf} - \varepsilon_\infty$, towards the value defined by the second term in the denominator of **eq. (10)**.

Finally, we give $\varepsilon''_{AP,0}$ and $\nu_0$ as a function of $1/\tau_D$ at different values of $\varepsilon_{lf} - \varepsilon_\infty$ (**Figure 7**). The value of $\varepsilon''_{AP,0}$ is practically independent of $1/\tau_D$, but the peak disappears at low $1/\tau_D$, because it is lost in the low frequency increase of $\varepsilon''_{AP}$ due to the finite resitance of the surface layers and the N$_F$ slab. The dependence $\nu_0(1/\tau_D)$ is linear at high enough values of $\tau_D$. When $\tau_D$ reduces and becomes smaller than $C_0 R_e \varepsilon_{lf}$, the value of $\nu_0$ is determined by the electrode resistance and not by the Debye relaxation time (see **eq. (10)**). From **Figure 7** it is also seen that the approximation given by **eq. (10)** is valid up to some maximum value of $\tau_D$. This value depends on the resistance of the surface layers or LC layer (whichever is higher). If at least one of this resistances is infinite, the dependence is linear (dotted lines in **Figure 7**).

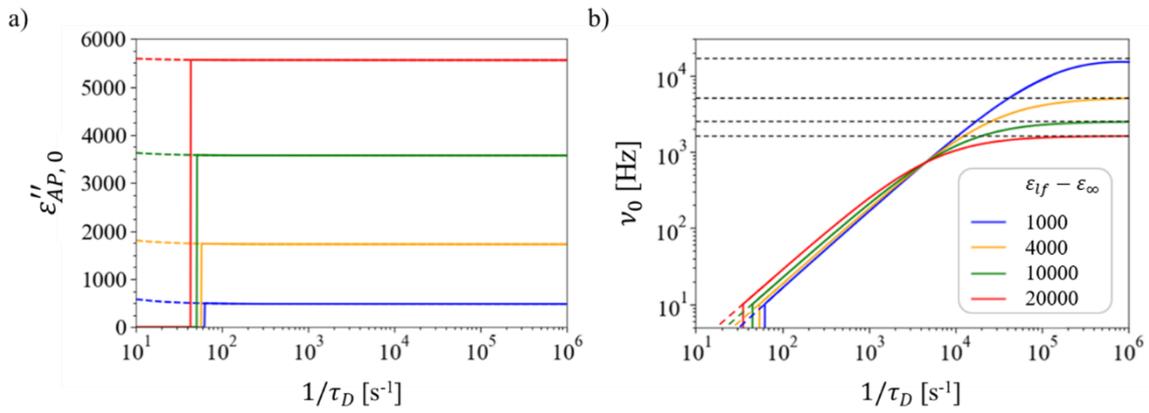

**Figure 7.** *High-$\varepsilon$ model.* a) $\varepsilon''_{AP,0}$ and b) $\nu_0$ as a function of $1/\tau_D$ at different values of $\varepsilon_{lf} - \varepsilon_\infty$. Black dotted lines in b) present $\nu_0$ at $\tau_D = 0$. Parameter values: $R_S = 1$ MΩ, $\rho_{LC} = 10^8$ Ωm (solid coloured lines), $\rho_{LC} = 10^9$ Ωm (dashed coloured lines), $R_e = 100$ Ω, $S = 100$ mm$^2$, $\varepsilon_S = 10$, $d_S = 2$ nm, $d = 10$ μm.

**Resistance of electrodes**

As shown above, the resistance of electrodes and other parasitic resistances, all included in $R_e$, can strongly influence interpretation of the DS results. Within the semi-quantitative fit of DS results presented in [18] we estimated that parasitic resistance was approximately 50 Ω for cells with gold electrodes, while in the case of ITO cells the resistance was of the order of few hundred Ω ($\approx 500$ Ω). A straightforward way to measure $R_e$ would be to find the frequency $\nu_0$ at which the real and imaginary part of complex impedance of an empty cell are equal. If we know the capacitance of an empty cell ($C_0$), we can calculate $R_e = (2\pi \nu_0 C_0)^{-1}$. However, frequencies $\nu_0$ are usually rather high and exceed the accessible range of standard impedance meters. The obvious solution is to connect a resistor with a known resistance in series with the cell. This will increase the $RC$ constant and reduce the frequency, but brings in additional problems, as the parasitic resistance is much lower than the resistance of the added resistor. At high frequencies, the resistance of the added resistor might also vary with frequency. Thus, results should be treated with caution. We estimated the electrode resistance by measuring the $RC$ constant of cells filled with the material in crystal phase at room temperature. We used cells with ITO electrodes from the previous research [18] (see **Table 1**). The measuring system presented in the Supplemental Material to [18] was optimised to increase accuracy at frequencies above 1 MHz. In addition, the whole electric circuit for the Bode Plot measurements (three cells measured at the same time) was designed on a high frequency board which significantly reduced parasitic resistances and mutual disturbances among the cell wiring due to the high frequency signals. The resistance of the resistor connected in series with the cell was $R_{add} = 200\ \Omega$. The resistance of the reference resistor (see Supplemental Material to [18]) was 5 k$\Omega$ and the amplitude of the voltage was 0.1 V. The frequency $\nu_0$ at which $\varepsilon''_{AP}$ is maximum (at this frequency the real and imaginary part of impedance are equal), see **Figure 8a**, is related to the $RC$ constant as $\nu_0 = [2\pi(R_{add} + R_e)C]^{-1}$. $C$ can be measured by a capacitance meter. However, from the Bode plot measurements one can also calculate the frequency dependent capacitance of the capacitor ($C$) and resistance of the resistor ($R = R_{add} + R_e$) that are connected in series (see **Figure 8b,c**). The capacitance at the frequency $\nu_0$ is always slightly smaller than the capacitance at lower frequencies. This difference was added as an error to the value of capacitance $C$ given in **Table 1**. The value of $R_e$, obtained from the value of $R$ at frequency $\nu_0$ is also given in **Table 1**. We see that for all cells with ITO electrodes, $R_e$ is of the order of few hundred Ohms. The obtained values are lower than the ones estimated in [18], which is expected, because the parasitic resistances were significantly reduced by the use of the high frequency board. Also, in [18], we searched only for a semi-quantitative agreement between the DS results and model predictions.

We also note that at room temperature the material filling the cell is in the solid state and the relative permittivity is low, so the apparent value given in **Figure 8a** can be considered as reflecting actual material properties. Taking twice the maximum value of $\varepsilon''_{AP}$, we find that $\varepsilon'$ of the material in the crystalline phase is 3.5, which is a reasonable value.

Finally, we observe that the capacitance of the cell as well as the resistance $R$ start to reduce at high frequencies. Because the peak value of $\varepsilon''_{AP}$ for all four cells is at frequencies where $R$ and $C$ are approximately frequency independent, we conclude, that the estimation of $R_e$ is valid.

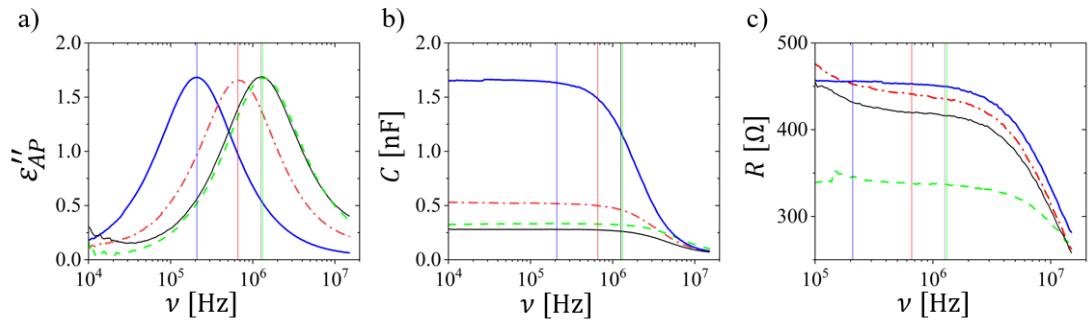

**Figure 8.** a) The frequency ($\nu$) dependence of the apparent value of the imaginary part of relative permittivity ($\varepsilon''_{AP}$) calculated from the Bode plot measurements of cells filled with the material at room temperature in a series connection with a resistor with resistance 200 $\Omega$. Frequency dependence of b) capacitance ($C$) and c) resistance ($R$) calculated from the Bode plot measurements. Cells: ITO electrodes with no surfactant layer: $d = 3.0\ \mu$m (solid blue curves), $d = 9.7\ \mu$m (dash-dotted red curves), $d = 4.9\ \mu$m (dashed greed curves) and ITO electrodes with surface layer of surfactant enforcing homeotropic anchoring: $d = 10.0\ \mu$m (thinner solid black curves). Vertical lines are drawn at frequencies of the peak value of $\varepsilon''_{AP}$.

**Table 1.** Measured parasitic resistances ($R_e$) for some of the cells used in the research reported in [18]. Electrodes: ITO, with no surfactant layer (NA) and with surface layer of surfactant enforcing homeotropic anchoring (HT). Cell thickness ($d$), capacitance of empty cells ($C_0$), room-temperature capacitance of cells filled with the N$_F$ material ($C$) and electrode surface area ($S$).

| cell | $d$ [μm] | $C_0$ [pF] | $S$ [mm$^2$] | $\nu_0$ [MHz] | $C$ [nF] | $R_e$ [kΩ] |
|---|---|---|---|---|---|---|
| NA | 3.0 | 485 | 164 | 0.209 | 1.63 ± 0.02 | 0.26 ± 0.01 |
|    | 4.9 | 95  | 53  | 1.32  | 0.32 ± 0.01 | 0.13 ± 0.02 |
|    | 9.7 | 150 | 164 | 0.661 | 0.52 ± 0.02 | 0.22 ± 0.02 |
| HT | 10.0 | 77 | 87  | 1.26  | 0.26 ± 0.02 | 0.22 ± 0.03 |

**Conclusions**

We have compared the dielectric response of a ferroelectric nematic phase as predicted by the PCG and high-$\varepsilon$ model. The PCG model predicts that all the cells with the same type of electrodes (and the same thickness of the thin layer of surfactant on the electrodes), filled with a N$_F$ material, will have the same capacitance independent of the cell thickness (see **eq. (11)**). On the other hand, the high-$\varepsilon$ model can account for the experimentally observed dependence of capacitance on the cell thickness (**eq. (12)**). The PCG model predicts that the frequency, up to which capacitors filled with N$_F$ effectively store energy, decreases with increasing cell thickness (**eq. (7)**), while the high-$\varepsilon$ model predicts that this frequency can strongly depend on the resistance of electrodes (**eq. (10)**) and can even increase with increasing the cell thickness. Both models predict that the frequency $\nu_0$ at which the apparent value of the imaginary part of the relative permittivity is maximum linearly increases with increasing the thickness of the thin layer of surfactant, if the thickness of this thin layer is high enough. However, at thicknesses below approximately 10 nm, there is a significant difference in prediction of both models (see **Figure 5**), as the high-$\varepsilon$ model predicts that $\nu_0$ does not reduce to 0. For both models it is shown that the predicted results strongly depend on the resistance of electrodes, see for example, the dependence of $\nu_0$ on the cell thickness in **Figure 4**. Especially in cells with ITO electrodes, the resistance of electrodes cannot be neglected, because we measured it to be 100 to 300 Ω.

Because the N$_F$ phase in a nonconductive phase and its response to external electric field is due to the rotation of permanent dipoles, we argue that the effect of rotation of dipoles results in a large relative permittivity (high-$\varepsilon$ model). Also, the PCG model cannot account for the thickness dependence of capacitance in cells with different thickness and the same type of electrodes and it predicts unphysically high values of the rotational viscosity, depending on the material also of the order of several hundred Pas. Thus, we find the high-$\varepsilon$ model as more appropriate to use for the description of the N$_F$ phase. However, the differences between the model predictions described in this paper can be used for further experimental investigation on the appropriateness of both models.

**Acknowledgements:** N.V. and V.M. acknowledge the support of the Slovenian Research and Innovation Agency (ARIS), through the research core funding Programs No. P1-0055 and No. P2-0368, respectively. E.G. and D.P. acknowledge the support by the National Science Centre (Poland) under Grant No. 2021/43/B/ST5/00240.